\documentclass{article}

\usepackage{arxiv}

\usepackage[utf8]{inputenc} 
\usepackage[T1]{fontenc}    
\usepackage{hyperref}       
\usepackage{url}            
\usepackage{booktabs}       
\usepackage{amsfonts}       
\usepackage{nicefrac}       
\usepackage{microtype}      
\usepackage{lipsum}		
\usepackage{graphicx}
\usepackage{natbib}
\usepackage{doi}

\usepackage{amsthm}
\usepackage{mathtools}
\usepackage{amsfonts,amssymb,amsbsy}
\usepackage{graphicx}
\usepackage{anysize}
\usepackage{amsmath}   
\usepackage{subcaption}
\usepackage{algorithm}
\usepackage{algorithmic}
\usepackage{ulem}

\title{Informative Bayesian model selection for RR Lyrae star classifiers}
\author{
\large
F. P\'erez-Galarce$^{1},$ \text{  }
 K. Pichara$^{1,2}$  \text{  } P. Huijse$^{3,2}$ \text{  } M. Catelan$^{2,4,5}$ \text{  } D. Mery$^{1}$
\\
$^{1}$Department of Computer Science, School of Engineering, Pontificia Universidad Cat\'olica de Chile, 7820436 Santiago, Chile \\
$^{2}$Millennium Institute of Astrophysics, Santiago, Chile\\
$^{3}$Instituto de Inform\'atica, Universidad Austral de Chile, Valdivia, Chile\\
$^{4}$Instituto de Astrof\'isica, Facultad de F\'isica, Pontificia Universidad Cat\'olica de Chile, Av. Vicu\~{n}a Mackenna 4860, 7820436 \\ Macul, Santiago, Chile\\
$^{5}$Centro de Astroingenier{\'{\i}}a, Pontificia Universidad Cat{\'{o}}lica de Chile, Av. Vicu\~{n}a Mackenna 4860, 7820436 Macul, Santiago, \\ Chile
}

\date{Accepted XXX. Received YYY; in original form ZZZ}

\begin{document}

\date{}

\maketitle

\begin{abstract}{

Machine learning has achieved an important role in the automatic classification of variable stars, and several classifiers have been proposed over the last decade. These classifiers have achieved impressive performance in several astronomical catalogues. However, some scientific articles have also shown that the training data therein contain multiple sources of bias. Hence, the performance of those classifiers on objects not belonging to the training data is uncertain, potentially resulting in the selection of incorrect models. Besides, it gives rise to the deployment of misleading classifiers. An example of the latter is the creation of open-source labelled catalogues with biased predictions. In this paper, we develop a method based on an informative marginal likelihood to evaluate variable star classifiers. We collect deterministic rules that are based on physical descriptors of RR Lyrae stars, and then, to mitigate the biases, we introduce those rules into the marginal likelihood estimation. We perform experiments with a set of Bayesian Logistic Regressions, which are trained to classify RR Lyraes, and we found that our method outperforms traditional non-informative cross-validation strategies, even when penalized models are assessed. Our methodology provides a more rigorous alternative to assess machine learning models using astronomical knowledge. From this approach, applications to other classes of variable stars and algorithmic improvements can be developed.     
      }
\end{abstract}

\section{Introduction}

\noindent Machine learning has been applied intensively to the classification of variable stars in recent decades \citep{debosscher2007automated,debosscher2009automated,richards2011machine,pichara2012improved,mackenzie2016clustering,benavente2017automatic,narayan2018machine,aguirre2018deep,nun2015fats,kim2016package,valenzuela2017unsupervised,naul2018recurrent,carrasco2018deep,becker2020scalable}. Variable stars are considered crucial celestial objects, mainly because several of them (e.g., RR Lyrae and Cepheids) are reliable distance indicators, thus providing us with a gauge to measure topics ranging from Galactic structure to the overall expansion of the Universe. 

From a machine learning perspective, the major efforts have been concentrated on developing variable star classifiers \citep{debosscher2007automated,debosscher2009automated,richards2011machine,pichara2012improved,mackenzie2016clustering,benavente2017automatic,carrasco2018deep,narayan2018machine,aguirre2018deep,becker2020scalable} and new alternatives to represent the celestial objects \citep[i.e., human-based and deep learning-based features;][]{ nun2015fats,kim2016package,valenzuela2017unsupervised,naul2018recurrent}. However, these classifiers rely heavily on the quality of the labelled training datasets and it is challenging and highly time-consuming to generate representative training datasets to train these models. This drawback hinders the model assessment procedure and, therefore, it can impact the performance of these models when they are tested on objects beyond the labelled objects (including those from the same catalogue). More significantly, this could lead to wrong models being used to label new training data, thereby generating a cascade effect. This situation has given rise to the following questions: \textit{Do we have confidence in our variable star classifiers? Can we improve the model assessment process?} To answer these questions, the metrics used to evaluate models under non-favourable conditions have become an important topic.  

Several papers have discussed the impact of biases in the training data of variable stars  \citep{debosscher2009automated,richards2012overcoming,masci2014automated}. However, to the best of our knowledge, no definitive solution has been proposed to more accurately assess the classifiers in this scenario. Bias in data means that there is a difference between the joint distribution of our labelled data $\mathcal{D}^{S}$ and that of the population $\mathcal{D}^{P}$, which considers all the observed objects for the astronomical project (survey). The problem arises from the fact that the current classifiers are trained with a subset of $\mathcal{D}^{S}$ and, subsequently, the performance is evaluated using the complement (the testing set), typically by means of a cross-validation (CV) scheme. The CV strategies assume the existence of representative training data; however, we know that datasets in astronomy are biased and, consequently, we are unable to report a realistic classification performance.  

Those biases stem from several sources, the majority of which can be linked to human-related tasks and technical characteristics of the telescopes. The former is associated with the labelling process since astronomers are more prone to label a class when it is easier to define. In this sense, \citet{cabrera2014systematic} presented a good discussion about this systematic bias in the astronomical labelling process. The mechanical design of receptors generates another type of bias, specifically in relation to the range in which the signal can be processed; for example, when distances increase, less luminous objects are more difficult to see \citep{richards2012overcoming}. Finally, the rapid development of technology accelerates the obsolescence of the models. Hence, we are unable to apply trained models to new surveys and we lack sufficient confidence in the error metrics in these newer catalogues. This problem is addressed by domain adaptation and was discussed in depth in variability surveys in \citet{benavente2017automatic}. 

To analyze the effect of these biases, they are typically divided into two categories: biases in features and biases in class representations. The existence of biases in features (e.g., period and amplitude) means that there is a difference in the joint feature distribution between $\mathcal{D}^{S}$ and $\mathcal{D}^{P}$. That is to say, zones of the feature space without labelled objects or an over-representation of other zones, it impacts the relevance of those zones during the training and assessment process. Bias in the representation of classes is associated with some classes of variable stars that are more/less represented in $\mathcal{D}^{S}$  compared to $\mathcal{D}^{P}$.

Notwithstanding this underlying problem, few efforts have been made to study metrics and validation strategies to evaluate the performance of light curve classifiers. Furthermore, it is a challenge to provide more accurate metrics to assess models in a scenario in which we cannot entirely trust the data. One natural framework with which to address the aforementioned problems is Bayesian modelling, which has been increasingly used in different fields of astronomy, such as to compare astrophysical models \citep{ford2006bayesian} or make predictions on the properties of celestial objects \citep{das2018made}. To improve the model assessment task, we propose a novel pipeline for evaluating Bayesian Logistic Regressions on biased training data. The methodology used is based on Bayesian machine learning, which allows us to incorporate astronomical knowledge into the model assessment process. Our approach exploits the powerful Bayesian Model Selection (BMS) scheme \citep{murray2005note}, which embodies desirable properties such as Bayesian Occam's razor, consistency, and comparability \citep{myung1997applying}. 

The BMS framework is based on the marginal likelihood (also known as Bayesian evidence), which is the likelihood function weighted by a prior distribution over the range of values for its parameters. In other words, the marginal likelihood contains the expected probability of data over the parameters.  However, if we do not add information to these prior distributions, even this powerful and robust metric is unable to assess the models correctly when the training data are biased. Hence, to address these biases, we contribute with a strategy that exploits expert knowledge by incorporating informative priors in the marginal likelihood estimation of RR Lyrae star classifiers. Our methodology is divided into three stages; first, we propose a method to represent the prior knowledge using deterministic rules (DRs) founded on physical-based features, such as period and amplitude. In the second stage, we generate posterior samples using these informative priors. This is a suitable approach since, by means of posterior samples, we are able to ensure zones of high value in the likelihood function and the prior distribution. Moreover, we can add astronomical knowledge through the effect of the priors in the posterior distribution. Finally, in the third phase, we estimate the marginal likelihood using an approximated sampling method. 

This paper is organized as follows. Section \ref{back} introduces the background theory of metrics and validation strategies for the assessment of models. Section \ref{rel} provides an account of related works and is divided into two subsections: firstly, we review machine learning models in the classification of variable stars which deal with biases; secondly, we present how Bayesian data analysis has been applied in the field of astronomy. Section \ref{marginallikehood} outlines the proposed methodology used to address the challenge.  Sections \ref{class} and \ref{implementation} describe the data and the implementation, respectively. After that, Section \ref{results} shows the results. Finally, Section \ref{label} sets out the conclusions and future work.    

\section{Background Theory}
\label{back}
In the machine learning and statistical learning fields, the model assessment process is a central topic and one which is typically associated with three main tasks: (i) the evaluation of a population error using the training data error; (ii) the selection of the most suitable model among a set of alternatives; and (iii) the definition of a good set of hyper-parameters. In this section, we summarize the traditional methods used to assess models as follows: Section \ref{metrics} concentrates on the metrics for model selection and Section \ref{evaluation} focuses on validation strategies.

\subsection{Metrics for evaluating classifiers}
\label{metrics}

There are several metrics to evaluate the performance of classification models which have been originated from different fields such as statistical learning, information theory and data mining. Consequently, selecting one metric or a set thereof to assess our models can become challenging. Given that, it is important to consider the following well-known basic properties when using or proposing a metric. Consistency: the size of the training data should not affect our metric; Occam's razor principle: we desire a metric that can identify whether a model has the optimal complexity required; comparison: it should allow us to compare non-nested models; reference: the metric must be independent of the validation strategy; and individuality: the metric must be able to measure any given object individually \citep{anderson2004model}. 

We present a summary of the most frequently used metrics below. They have been divided into two groups;  Section \ref{confusionmatrix} presents metrics based on the confusion matrix and Section \ref{bayesianModelSelection} provides a scheme of methods based on BMS. 

 \subsubsection{Metrics based on confusion matrix}
\label{confusionmatrix}

Within this framework, the most intuitive metric is the Accuracy, which evaluates prediction quality based on the ratio of correct predictions over the total number of observations. This metric has two critical drawbacks: firstly, it is not able to discriminate the type of error, and secondly, it can be easily dominated by the majority class.

In order to assess the type of error, other measures can be obtained. For example, the Recall, which represents the fraction of positive patterns that are correctly classified. Or the Precision, which corresponds to the ratio between the positive objects that are correctly predicted and the total number of predicted objects for the true class. To consider a balance between Recall and Precision, we can evaluate these two metrics in conjunction through the F1-score. This metric is the harmonic-mean between Precision and Recall, and it is more robust than Accuracy when the dataset has imbalanced classes. 

The aforementioned metrics are the most common ones in this framework, although there are many variants in the literature. A summary of these can be found in \citet{sokolova2009systematic}. Despite the large variety of metrics, there are a number of related limitations: (i)
We cannot compare the trade-off between the goodness-of-fit and the model complexity directly.
(ii) We must use validation strategies. 
(iii) Due to the fact that these metrics consider a hard classification (i.e., a Boolean decision about the predicted class), we cannot consider different levels of confidence in the prediction scores.

\subsubsection{Bayesian Model Selection}
\label{bayesianModelSelection}

A robust alternative for selecting models is the marginal likelihood, which is denoted by $p(\mathcal{D}|m)$, where $m$ represents a model and $\mathcal{D}$ the training data. It appears in the first level of inference in the Bayesian framework, \begin{align}
 p(\theta| \mathcal{D}, m )& = \frac{p(\mathcal{D}|\theta, m )p(\theta|m )}{p(\mathcal{D}| m)}, \label{bayesianframe} \\
 p(\mathcal{D} |  m ) & = \int p(\mathcal{D}|\theta,  m )p(\theta| m)d\theta.  \label{marglik}
\end{align}

We use the following traditional notation: let  $p(\theta| \mathcal{D}, m )$ be the posterior distribution of the parameter given the data and a model; let $p(\mathcal{D}|\theta, m )$ denote the likelihood function; let $p(\theta|m )$ represent the prior distribution over the parameters and finally, let $p(\mathcal{D}| m)$ be the marginal likelihood.

The marginal likelihood like a model selector was analyzed  in depth by \citet{mackay1992bayesian} and, subsequently,  the links with the Occam's razor principle were emphasized in  \citet{rasmussen2001occam}, \citet{murray2005note}, and \citet{ghahramani2013bayesian}. The idea of using the marginal likelihood in model assessment comes from the second level of inference, 
\begin{align}
        p(m|\mathcal{D})= \frac{p(\mathcal{D}|m)p(m)}{\sum_{m \in \mathcal{M}}p(m, \mathcal{D})}, \label{modelselection}
\end{align}

\noindent where the Bayes' theorem is used to estimate the model probability given a dataset $p(\mathcal{D}|m)$.

The estimation of $p(m|\mathcal{D})$ is intractable since we cannot enumerate all possible models. However, we can apply the same criteria used in the first level of inference, avoiding the denominator estimation (constant). In this way, we can estimate the model posterior by $p(m|\mathcal{D}) \propto p(\mathcal{D}|m)p(m)$. Finally, if we assume a non-informative prior for the model, $p(m|\mathcal{D})$ is proportional to the marginal likelihood. For this reason, we use the marginal likelihood to select the most appropriate model. 

Despite the fact that the marginal likelihood automatically embodies all those desired properties of good metrics, it is unable to automatically manage the biases in the training data and its estimation is a computational challenge in high-dimensional data (see equation \ref{marglik}). To address this challenge, we can estimate the marginal likelihood by interpreting it as an expected value and then performing Monte Carlo (MC) estimation according to the following equations: 
 
\begin{align}
        p(\mathcal{D}|m)& = \mathbb{E}_{\theta} \left[ p(\mathcal{D}| \theta, m)\right],         \label{expectedvalue} \\
       & \frac{1}{S}\sum_{s=1}^{N}p(\mathcal{D} \mid \theta_s, m),  \theta_s \sim p(\theta). \label{eq:MaxE}
\end{align}

This simple approach only performs well if the prior and likelihood have a similar shape and are strongly overlapped. If this does not hold, then misleading samples can be generated in low-valued areas of the likelihood function. Therefore, a few samples with high values in the likelihood function dominate the estimator, and this could produce a high variance in the estimation procedure \citep{gronau2017tutorial}.

Due to these difficulties, the majority of research into BMS avoids MC sampling methods by applying approximations such as the Laplace approximation and BIC (Bayesian Information Criterion) \citep{schwarz1978estimating,watanabe2013widely} or they resort to MC methods based on posterior samples \citep{neal2001annealed, raftery2006estimating, overstall2010default}. Our proposal is based on the latter type of strategies but adding astronomical knowledge to those posterior samples.   

\subsection{Validation strategies} 
\label{evaluation}

CV is the most common family of methods for estimating metrics. We present a summary and some drawbacks of the three most common CV-based methods. Firstly, we review \textit{Hold-out}, which is the most basic approach. Therein, two sets of data are generated, one of which is used to train the model and the other to evaluate its quality. This approach depends heavily on one dataset, and for that reason, it is a good option only when we are in possession of large quantities of data or when there is some running time limitation. Moreover, in small datasets, hold-out can generate a pessimistic estimator \citep{lendasse2003model}. 

Secondly,  \textit{k}-fold, which is the most commonly used variant of CV, considers splitting the data $\mathcal{D}^{S}$ into smaller chunks $\mathcal{D}_1^{S}, \mathcal{D}_2^{S}, ... \mathcal{D}_K^{S}$ with the same size. We train using $\mathcal{D}^{S} \setminus \mathcal{D}^{S}_k$ chunks and evaluate using the free chunk $\mathcal{D}^{S}_k$. The number of folds provides the bias-variance trade-off; a small number of folds reduces the bias but increases the variance. Lastly, \textit{Leave-One-Out} uses each data point as a chunk, and for each object, a model is trained to leave only this object out. It provides an unbiased estimator, although its variance can be larger, and the running time can be prohibitive \citep{rao2008dangers}.  

\citet{arlot2010survey} presented a survey of CV procedures for model selection. Despite the effort to develop variants, these ideas consist of a fundamental assumption. They consider that we are working with representative training data. This means that $\mathcal{D^{S}}$ has the same probability distribution as the data beyond the labelled objects $\mathcal{D^{T}}$. However, in astronomy, we are often unable to generate such representative training data.

In an attempt to overcome this challenge, \citet{sugiyama2007covariate} proposed a CV variant to tackle the aforementioned biases (also known as \textit{data shift} problem) by means of an importance weighted CV  (IWCV) approach. The IWCV weighs each observation $i$ in the evaluation metric with the density ratio $p(x_i)_{\text{test}} \mathbf{ \mathbin{/}} p(x_i)_\text{train}$.  Note that IWCV is proposed to address a type of \textit{data shift}, which is named  as a covariate shift (bias in features), here, $p(x_i)_{\text{train}}\neq p(x_i)_{\text{test}}$, but $p(y|x)_{\text{train}} = p(y|x)_{\text{test}}$. If we need to work on scenarios with bias in labels (target shift), which assumes  $p(y)_{\text{train}}\neq p(y)_{\text{test}}$, but $p(x|y)_{\text{train}} = p(x|y)_{\text{test}}$, we must adapt the density ratio by $p(y)_{\text{train}}\mathbf{ \mathbin{/}}p(y)_{\text{test}}$. This is a clever approach, but it assumes an existing knowledge of the probability density functions for  $\mathcal{D}^{S}$ and  $\mathcal{D}^{T}$, which can be intractable in high dimensions. 

A further commonly used method is bootstrap \citep{efron1997improvements},  which uses sampling with replacement, in which $N$ samples are selected in each of the $k$ iterations. In this approach, in a given iteration, a particular sample may appear more than once, while others might not appear at all. Although the traditional bootstrap has interesting statistical properties, it fails to select classifiers in a machine learning context because it favours overfitting classifiers \citep{kohavi1995study}. 

 An interesting variant is found when the bootstrap approach is analyzed from a Bayesian perspective \citep{rubin1981bayesian}. A traditional bootstrap can be understood by modelling the probability of drawing a specific observation such as a categorical distribution, \textit{Cat($\pi$)}, where the vector $\pi=(\pi_1, \pi_2, ..., \pi_N)$ is the probability of drawing each object ($\sum_{i}^{N}\pi_i = 1$). In a traditional bootstrap we have $\pi_1 = \pi_2 = \pi_k = \pi_N = 1\mathbf{ \mathbin{/}}N$. In a Bayesian view, $\pi$ draws from a Dirichlet, \textit{Dir($\alpha$)}, where for example, the expected proportion for $\pi_1$ is based on the priors $\alpha_1\mathbf{ \mathbin{/}} \sum_{i=1}^{N}\alpha_i$. However, the generation of informative priors $Dir(\alpha)$ can pose a significant challenge.

\section{Related work}
\label{rel}

\noindent This section is divided into two subsections.  Section \ref{modelselectionrev} studies the state of the art of variable star classifiers, with emphasis on research addressing underlying biases and how these approaches select and compare models. Section \ref{BDA} discusses briefly how Bayesian data analysis has been used in the field of astronomy. 

\subsection{Classification of Variable Stars under bias}
\label{modelselectionrev}
\noindent  Several papers on the automatic classification of variable stars have sought to address the data shift problem from different perspectives. However, none has focused on model selection strategies. Over a decade ago,  \citet{richards2011active}  proposed several strategies to improve the training data. For example, they designed an active learning method and presented an importance-weighted CV method to avoid under-represented zones of feature space. However, metrics to compare models in these contexts were not analyzed in depth. \citet{masci2014automated}  proposed a random forest (RF) classifier, which was trained with a labelled set from the Wide-field Infrared Survey Explorer \citep[WISE;][]{wright2010wide}, within an active learning approach. This classifier was able to improve the training data and mitigate the biases. This RF approach outperformed Support Vector Machine (SVM), K Nearest Neighbours (KNN) and Neural Networks (NN) using a CV method to estimate the Accuracy.  

\citet{benavente2017automatic} proposed a full probabilistic model to address the domain adaptation problem. This model was able to transfer knowledge (feature vectors) among different catalogues. It was able to manage the covariate shift and improve the cross-validated F1-score. A Gaussian Mixture Model representing each catalogue (source and target) and a mixture of linear transformations (translation, scaling, and rotation) were applied. Recently, \citet{aguirre2018deep} designed a convolutional NN that was able to learn from multiple catalogues, outperforming an RF based on handcrafted features. To manage the imbalanced classes, \citet{aguirre2018deep} proposed a novel data augmentation scheme which creates new light curves by modifying real objects. 

\citet{sooknunan2018classification}  reported the relevance of a non-representative $\mathcal{D}^{S}$ when applying trained models on data from new telescopes. Moreover, they studied how the Accuracy metric decreases (training vs real) when $\mathcal{D}^{S}$ is small. To create the training data, they used a few real objects and synthetic light curves generated using a Gaussian Process. Experiments with the following five classes of transients were conducted: active galactic nuclei (AGN), supernovae (SNe), X-ray binaries (XRBs), $\gamma$-ray bursts (GRBs), and novae. The results led to the conclusion that a better performance can be obtained in new surveys if contextual information (object location) and multi-wavelength information are incorporated. To encourage the use of multi-wavelength information, they presented results using both the optical telescope MeerLICHT \citep{bloemen2016meerlicht} and the radio telescope MeerKAT \citep{booth2012overview}. 

\citet{naul2018recurrent} proposed the use of a Recurrent Autoencoder to learn a variable star embedding. The measurement error in observations is used for weighting the reconstruction metric in the loss function so that those observations with large measurement error were less important. Subsequently, this embedding is used to classify by means of an RF classifier. The new representation is compared with two baseline sets of handcrafted features \citep{kim2016package,richards2011machine}, being competitive with traditional approaches when folded light curves were used. It outperformed or was similar to the baselines in the LIncoln Near-Earth Asteroid Research (LINEAR) survey \citep{sesar2013exploring}  and the MAssive Compact Halo Object (MACHO) catalogue \citep{alcock1997macho}.

Recently,  \citet{becker2020scalable} presented a scalable Recurrent NN which was capable of learning a representation without human support. The researchers obtained a competitive Accuracy in shorter running time than an RF that was based on handcrafted features. Furthermore, they provided a comparison between biases affecting handcrafted features and those based on deep-learning features, thereby supporting the line of thought that deep learning models are capable of learning features that are less biased when working in specific surveys.

Table \ref{litrevtab} provides a summary of model assessment strategies for variable star classifiers. We conclude that the majority of the papers analyzed herein have applied metrics based on the confusion matrix and have primarily utilized $k$-fold for the validation thereof.

\begin{table*}
\begin{minipage}{160mm}
\footnotesize
\centering
\caption{Summary of strategies for selecting variable star classifiers. The bold letters in the Classifiers column represent  the best model according to the papers listed in the final column. GMM = Gaussian Mixture Model Classifier, BN = Bayesian Network, BAANN = Bayesian Average of Artificial Neural Networks, SVM = Support Vector Machine, CNN = Convolutional Neural Network, RNN = Recurrent Neural Network}
\label{litrevtab}
\begin{tabular}{cccc}
\hline
 Classifiers & Metrics  &  Validation & \textbf{Reference}  \\
  \hline
GMMC, BN, \textbf{BAANN}, SVM & Accuracy &  10-fold&  \citep{debosscher2007automated,debosscher2009automated}\\ 
  \hline
 CART, \textbf{Random Forest},  & Error rate & 10-fold &  \citep{richards2011machine}\\
  Boosted Trees, C4.5, SVM &  &  & \\
  \hline
RF & Error rate & $k$-fold &  \citep{bloom2012automating} \\
  \hline
\textbf{Boosted RF}, regular RF, SVM & F1-score & 10-fold &  \citep{pichara2012improved}\\ 

\hline 
     RF+BN& F1-score & $k$-fold& \citep{nun2014supervised}\\ 
 \hline
 NN, \textbf{RF}, SVM, KNN & Accuracy, ROC& Hold-out  & \citep{masci2014automated}\\ 
 \hline
 
 Meta Classifier (RF) & Precision-F1-score-Recall &  $k$-fold & \citep{pichara2016meta} \\ 
 \hline 
 \textbf{SVM}, RF & F1-score & 10-fold  & \citep{mackenzie2016clustering}\\
 \hline
  LR, RF, CART, \textbf{SBoost}, \textbf{AdaBoost} & Precision, Recall, F1, AUC & 10-fold & \citep{elorrieta2016machine}\\
   SVM, LASSO, NN, DNN &  & & \\
 \hline
 \textbf{RF}, SVM & F1-score & CV& \citep{benavente2017automatic}\\ 
 \hline
 Decision Tree & F1-score & bootstrap & \citep{castro2017uncertain}\\
 \hline
 Recurrent CNN, RF & Accuracy, Av. Recall& Hold-out& \citep{carrasco2018deep}\\
 \hline 
 RF & Accuracy & 3-fold & \citep{sooknunan2018classification}\\ 
  \hline
  CNN & Recall, F1-score, MC  & repetitive Hold-out & \citep{mahabal2017deep}\\
  \hline
 RF & OOB - Accuracy- ROC  & $k$-fold & \citep{narayan2018machine}\\
 \hline
 AE-RNN+RF & Accuracy & 5-fold & \citep{naul2018recurrent} \\ 
 \hline
 CNN, RF & Accuracy & 10-fold & \citep{aguirre2018deep} \\ 
 \hline
 \end{tabular}
 \end{minipage}
\end{table*}

\subsection{Bayesian Data Analysis in Astronomy}
\label{BDA}

\noindent In recent decades, several astronomical papers have proposed the application of a Bayesian analysis. For example, pioneering research was conducted by \citet{gregory1992new} and \citet{saha1994unfolding} on the parameter estimation of astrophysical models. The research field most heavily influenced by these developments has been probably that of cosmological parameter estimation \citep{christensen1998markov,christensen2001bayesian}. Accordingly, \citet{trotta2008bayes} provided a comprehensive review of Bayesian statistics with an emphasis on cosmology.  \citet{sharma2017markov} produced a literature review that focuses on the Monte Carlo Markov Chain (MCMC) for Bayesian analysis in astronomy, providing an extensive overview of several MCMC methods, while also emphasizing how astronomers have used Bayesian data analyses in the past and how such approaches should, in fact, be used more commonly in the present. Furthermore, \citet{sharma2017markov} exemplified a number of basic concepts for model selection in a Bayesian approach. Subsequently, \citet{hogg2018data} provided a pedagogical overview of MCMC in astronomical contexts and discussed its foundations, highlighting certain aspects to consider to avoid obtaining misleading results from applications of this otherwise powerful technique. Moreover, several papers have shown the advantages of the Bayesian model selection approach \citep{parviainen2013secondary,ruffio2018bayesian} in astrophysical model selection. 

\citet{weinberg2013computational} presented a software package to apply Bayesian statistics in astronomy, including methods for estimating the posterior distribution and managing the model selection. This paper also provides a comprehensive introduction to Bayesian inference. Moreover, \citet{weinberg2013computational} included two applications where the system performance on astrophysical models (Semi-analytic galaxy formation model and Galaxy photometric attributes) is evidenced. 

\citet{budavari2017faint} designed an incremental Bayesian method to decide whether observations correspond to faint objects or noise from the data set (multi-epoch data collection). To classify each object, thought a Bayes factor scheme the marginal likelihoods of competing hypotheses (object or no object), at each epoch, are compared. In order to define these hypotheses, expert knowledge of the flux of each alternative is included. 

In spite of the fact that several papers have applied BMS to astronomy, to the best of our knowledge, our proposal is the first approach that adds physical information during the assessment process of machine learning classifiers for variable stars.

\section{Informative Bayesian Model Selection}
\label{marginallikehood}

This section provides a comprehensive description of our method to add human knowledge to the assessment and selection of RR Lyrae star classifiers. The methodology assumes that we have a set of models $\{m_1, m_2,.., m_i, .. m_n\} \in \mathcal{M}$ and a biased set of labelled objects (variable stars) to train them. Our goal is to rank these models to obtain a good performance in a shifted dataset (testing set).  

The method can be divided into three main steps. Section \ref{priors} focuses on obtaining priors from DRs. Section \ref{posterior} considers the generation of posterior samples running an MCMC algorithm. Section \ref{marginallikelihood} presents the mechanism to add informative posterior samples to the marginal likelihood estimation procedure. 

\begin{figure*}
  \centering
  \includegraphics[width=0.75\linewidth]{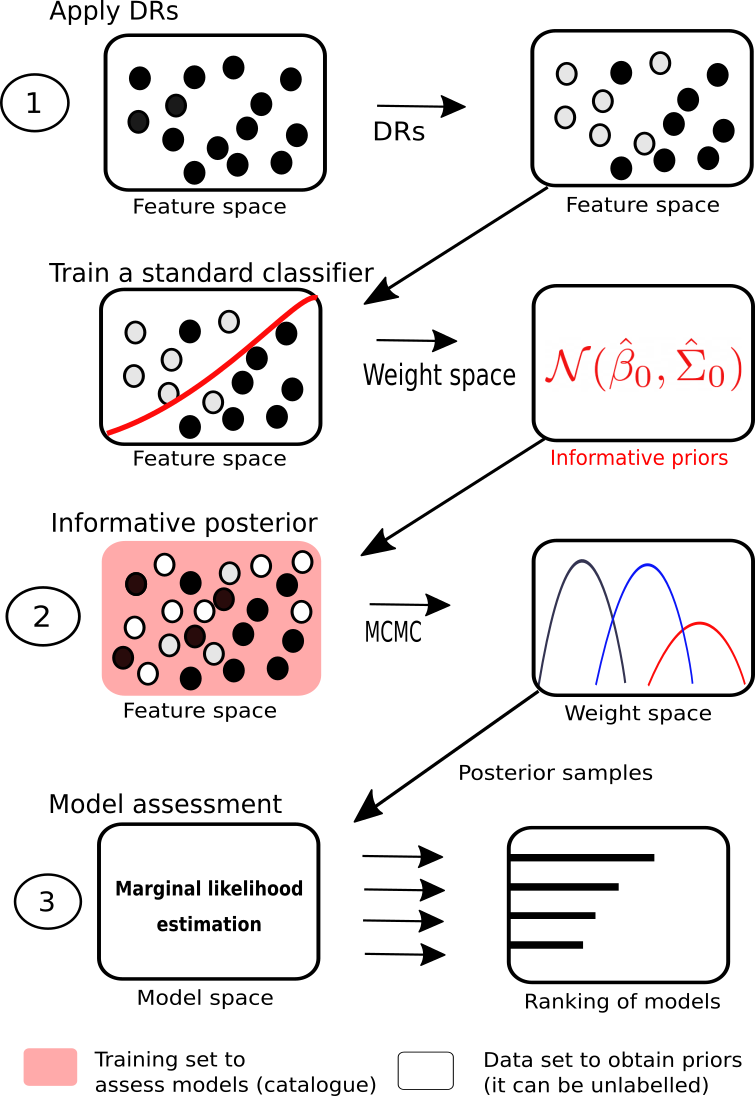}  
  \caption{ Proposed method overview. 1) First we label a variable star data set using the DRs. 2) Then, we train a standard logistic regression to obtain its weights. 3) After that, the mean and variance of those weights are used in an MCMC frame to generate posterior samples. 4) The marginal likelihood is estimated by using these informative posterior samples. Lastly, the estimated marginal likelihoods are used to rank the models.  }
  \label{procedure}
 \end{figure*}

Figure  \ref{procedure} shows a diagram of our method, in which the output for each step is highlighted. The final output is a ranking of models based on an informative estimation of the marginal likelihood. We propose to mitigate biases through this informative marginal likelihood.  

\subsection{Obtaining informative priors}
\label{priors}

In the Bayesian framework, informative priors offer a great opportunity to add expert knowledge to machine learning models; however, the majority of Bayesian approaches use non-informative priors, and hence, they rely completely on the likelihood function \citep{gelman2017prior}. The use of non-informative priors, if there is expert knowledge, can be controversial \citep{gelman2008weakly,golchi2019informative}, and it is valid for both levels of inference: the first level, when we make inference on parameters, and the second level, when we make inference on models. 

For some models, the addition of human knowledge can be less complex, since it can be transferred from the space of features to the space of parameters directly; this is the case of Bayesian GMM or Bayesian Naive Bayes. However, in models like Bayesian Logistic Regressions (BLR) or Bayesian Neuronal Networks, it is not direct. Proposing informative priors for BLR can be a great challenge \citep{hanson2014informative}, despite the fact that some alternatives have been proposed to add expert knowledge when it is available. \citet{gelman2008weakly} proposed weakly informative priors (Cauchy priors) that are also useful to solve the complete separation problem \citep{zorn2005solution}. \citet{hanson2014informative} provided an informative $g$-priors approach; this scheme is suitable if there is information about the probability of each class. In spite of these proposals, and to the best of our knowledge, information about the relationship between classes and features cannot easily be incorporated. To face this challenge, we propose a novel methodology to obtain informative Gaussian priors for BLR classifiers.      

We propose to obtain astronomical knowledge through DRs. DRs can be used to filter celestial objects without resorting to machine learning methods. The DRs are based on physical features such as period, mean magnitude, and amplitude. To design these rules for RR Lyrae stars, we can make use of literature in the field to define physical features that may be particularly relevant in characterizing this class of variable stars.

DRs can be understood as a relationship between an antecedent (if) and a consequent (then). To define a rule, we use a standard notation, $A\Rightarrow B$, where $A$ represents a physical condition (antecedent) and $B$ represents a class of variable stars (consequent). Some examples of DRs for pulsating stars include:

\begin{itemize}
    \item (period $\in [0.2,1.0]$ days)  $\Rightarrow$ RR Lyrae
    \item  (amplitude $\in [0.3-1.2]$ in $V$-band)  $\Rightarrow$ RR Lyrae
    \item (amplitude $\in [0.2-0.8]$ $I$-band) $\Rightarrow$ RR Lyrae
    \item (period $\in [1,100]$ days)  $\Rightarrow$ Classical Cepheid
    \item (period $\in [0.75,30]$ days)  $\Rightarrow$ Type II Cepheid
    \item (period $\in [0.5,8.0]$ hours)  $\Rightarrow$ Dwarf  Cepheid
\end{itemize}

Note that some physical conditions can be valid for more than one variable star class; however, when applying a chain of several DRs, this drawback is reduced. Despite that, we recommend mitigating this possible overlap by using not only various DRs but also DRs based on features that do not vary across different surveys (invariant features), e.g., period and amplitude \citep{catelan2014pulsating}. 

Once we have obtained the DRs, we propose Algorithm~1 to obtain informative priors. This algorithm identifies priors in a binary classification scheme; thus, we must use a set of rules for each class of variable stars.  

The priors $\hat{\theta}$ are generated by fitting a standard (non-Bayesian) Logistic Regression. The training data with which to fit this model becomes critical at this stage. This because depending on the training set,  our DRs can find a different distribution of objects for both the true class and the false class. It is possible to use an entire survey, a subset of a survey, or even an improved set (data augmentation, adversarial examples, down-sampling or over-sampling).     

This method allows transferring astronomical knowledge from the space of physical features to the space of model parameters through the collected DRs based on physical features of RR Lyrae stars. In particular, we define the mean estimator vector, $\mathbf{\hat{\theta}}$, and the variance estimators, Var($\hat{\theta}$),  for a normal prior. Var($\hat{\theta}$) is defined by the diagonal of the inverse Fisher Information matrix $\mathbf{I(\theta)}$. To avoid very small values for the estimated prior of variance Var($\hat{\theta}$), we add a small constant $\epsilon$ (for example $\epsilon = $0.1) after applying Algorithm~1. 
  
\begin{center}
\begin{minipage}{.8\linewidth}
\begin{algorithm}[H]\small   
\begin{algorithmic}
\renewcommand{\algorithmicrequire}{\textbf{Input:}}
\renewcommand{\algorithmicensure}{\textbf{Output:}}
\REQUIRE Data $\mathcal{D_X}$, classifier $m$, DRs 
\ENSURE weights for the classifier $m$, $\mathbf{\beta}$
\STATE 
$\mathcal{D_Y}=\mathbf{1}$
\FOR{$r \in$ Rules}
    \FOR{$d \in \mathcal{D_X}$}
    \STATE state$  \leftarrow$ r.applyDR($d$) 
        \IF {state $==$ False}
            \STATE $\mathcal{D_Y}$[$d$] = 0
        \ENDIF 
    \ENDFOR
\ENDFOR

\STATE $\mathbf{\hat{\theta}}  \leftarrow$  m.fit$(\mathcal{D_X},\mathcal{D_Y})$
\STATE Var$\mathbf{(\hat{\theta})}  \leftarrow$  diag( $\mathbf{I(\theta)}^{-1}$) 
\RETURN $\hat{\theta},$ Var$\mathbf{(\hat{\theta})}$
\end{algorithmic}
\caption{Procedure to obtain priors from physical based features}
\end{algorithm}
\end{minipage}
\end{center}

\subsection{Posterior samples generation}
\label{posterior}

Our path for transferring human knowledge is by means of posterior samples since these contain both prior knowledge and data information. In this step, for each $m \in \mathcal{M}$, we train an BLR with priors obtained using Algorithm~1. 

To estimate the posterior $p(\theta | \mathcal{D_X}, m)$, we propose to use standard MCMC techniques, such as Metropolis-Hastings or Hamiltonian Monte Carlo algorithms. Moreover, we use the Gelman-Rubin test to validate the sample convergence in each dimension \citep{gelman1992inference}. Lastly, to manage imbalanced classes, we downsample the datasets. 

This step is time-consuming; hence, more efficient sampling strategies could speed up our strategy. We did not consider Variational Inference since the samples from this approach can be biased and our approach requires precise and unbiased samples \citep{blei2017variational}.  

\subsection{Informative marginal likelihood estimation}
\label{marginallikelihood}

The marginal likelihood has been widely studied to compare and select machine learning models, despite the fact that its estimation represents a significant computational challenge. Comprehensive references for the study of estimation methods can be found in \citet{gronau2017tutorial} and \citet{wang2018new}. 

We propose addressing an informative estimation of the marginal likelihood by using a bridge sampling approach \citep{overstall2010default,gronau2017tutorial}. Unlike standard Monte Carlo estimators (Importance Sampling or Harmonic Mean Estimator), Bridge Sampling allows us to avoid dealing with typical constraints of standard Monte Carlo methods in relation to the shape of a proposal probability distributions. Indeed, this method has suitable properties in our context, mainly due to the following reasons: (i) it does not waste resources by generating samples in low-value zones,  and (ii) it allows us to incorporate astronomical knowledge in order to reduce the impact of biases in the training data. 

The Bridge Sampling estimator is based on a ratio of two expected values as follows:

\begin{equation}
\small 
p(\mathcal{D}) = \frac{\mathbb{E}_{g(\theta)}\left[  p(\mathcal{D|\theta})p(\theta)h(\theta)  \right]}{ \mathbb{E}_{p(\theta | \mathcal{D})} \left[  h(\theta)g(\theta)    \right]}.\label{mlikelihood}
\end{equation}

To estimate $\mathbb{E}_{g(\theta)}\left[  p(\mathcal{D|\theta})p(\theta)h(\theta)  \right]$, we use samples from a proposal distribution, $g(\theta)$, and to estimate $\mathbb{E}_{p(\theta | \mathcal{D})} \left[  h(\theta)g(\theta)    \right]$ we need posterior samples, $p(\theta | \mathcal{D})$, which contain astronomical knowledge.

The desired match between the samples from the proposal and those from the posterior is managed through a  function, which is named bridge function, 

\begin{equation}
\label{bridgeFunction}
h(\theta) = C\frac{1}{s_1 p(\mathcal{D}|\theta)p(\theta) + s_2p(\mathcal{D})g(\theta)}, 
\end{equation}

\noindent which plays a central role in the Bridge Sampling Estimator \citep{meng1996simulating}. When the bridge function is introduced to the estimator, the function depends recursively on $p(\mathcal{D})$; hence, for estimating it, it is solved iteratively by

\begin{align}
\hat{p}(\mathcal{D})^{t+1} &                                       = \frac{\frac{1}{N_2} \sum_{i = 1}^{N_2}\frac{p(\mathcal{D}| \theta_i) p(\theta_i)}{s_1 p(\mathcal{D}|\theta_i)p(\theta_i) + s_2\hat{p}(\mathcal{D})^{t}g_(\theta_i) }}{\frac{1}{N_1} \sum_{j = 1}^{N_1}\frac{g(\theta_j)}{s_1 p(\mathcal{D}|\theta_j)p(\theta_j) + s_2\hat{p}(\mathcal{D})^{t}g_(\theta_j) }}, \label{IterativeBridge}\\                            
& \theta_j \sim p(\theta|\mathcal{D});  \theta_i \sim g(\theta). \nonumber
\end{align}

Through this estimator, astronomical knowledge is incorporated into the assessment process. By using an informative prior, we can reduce the effect of biases in the training sets on the posterior.  A proof of this estimator is presented in Appendix \ref{bridgeSampling} \citep{gronau2017tutorial}.

\begin{table*}
\centering
\caption{Class distribution of OGLE labelled set.}
\begin{tabular}{lcc}
\hline
 Class & Abbreviation & Number of objects  \\
  \hline
Long-Period Variable & lpv &     323,999\\
RR Lyrae & rrlyr &    42,751\\
Eclipsing Binary & ecl    &   41,787\\
Cepheids & cep &       7,952\\
Delta Scuti  & dsct &      2,807\\
Type II Cepheid & t2cep &      589\\
Double Periodic Variable & dpv    &     135\\
Anomalous Cepheid & acep    &     81\\
Dwarf Nova & dn       &    35\\
R CrB Variable & rcb      &    22\\
  \hline
\end{tabular}
\label{table1}
\end{table*}
\vspace{2pt}

\section{Data and Classifiers} 
\label{class}
\begin{figure*}
\begin{minipage}{165mm}
\begin{subfigure}{.5\textwidth}
  \centering
  \includegraphics[width=1\linewidth]{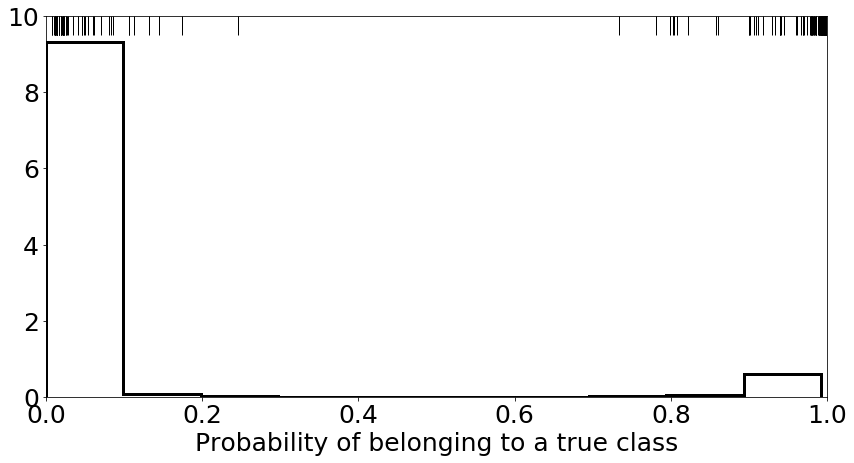}
  \caption{}
  \label{fig:sfig1}
\end{subfigure}%
\begin{subfigure}{.5\textwidth}
  \centering
  \includegraphics[width=1\linewidth]{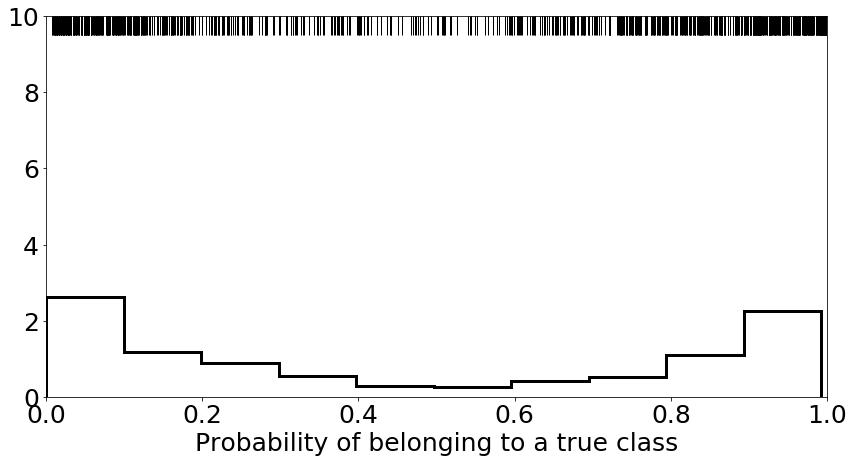}
  \caption{}
  \label{fig:sfig2}
\end{subfigure}
\begin{subfigure}{.5\textwidth}
  \centering
  \includegraphics[width=1\linewidth]{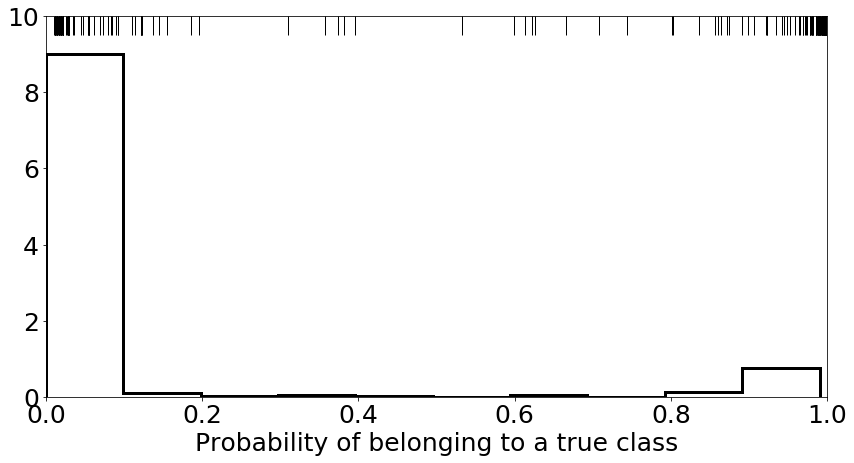}
  \caption{}
  \label{fig:sfig6}
\end{subfigure}%
\begin{subfigure}{.5\textwidth}
  \centering
  \includegraphics[width=1\linewidth]{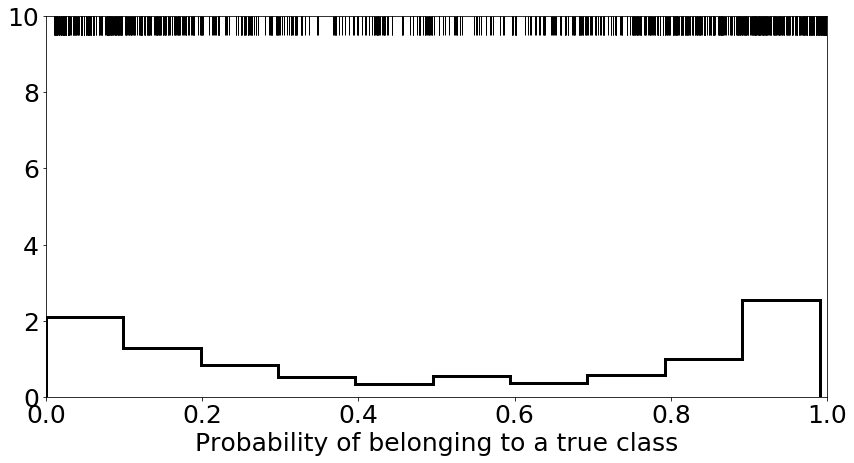}
  \caption{}
  \label{fig:sfig3}
\end{subfigure}
\begin{subfigure}{.5\textwidth}
  \centering
  \includegraphics[width=1\linewidth]{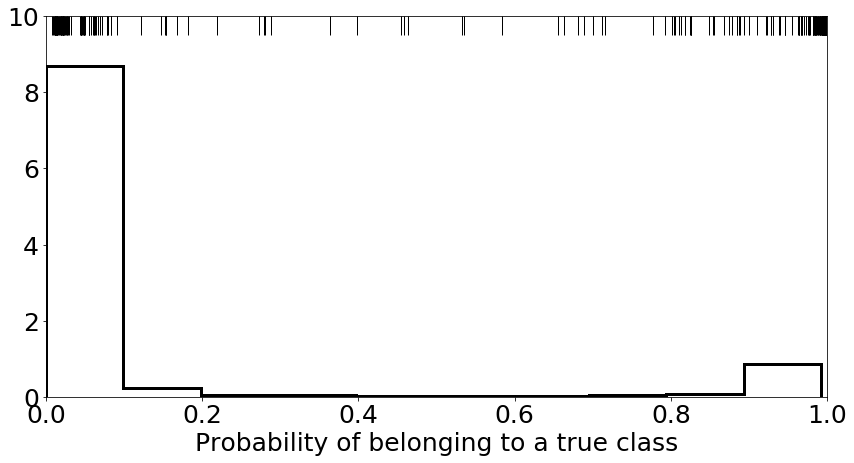}
  \caption{}
  \label{fig:sfig4}
\end{subfigure}%
\begin{subfigure}{.5\textwidth}
  \centering
  \includegraphics[width=1\linewidth]{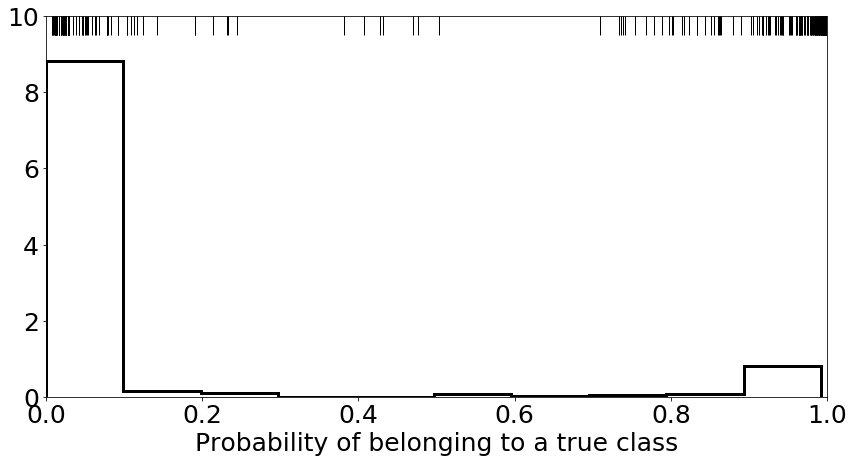}
  \caption{}
  \label{fig:sfig5}
\end{subfigure}
\caption{Histogram for the probability of belonging to the true class in rrlyrae-1 (a-b), rrlyrae-2 (c-d) and rrlyrae-3 (e-f). (a), (c) and (d) are training sets and (b), (e) and (f) are testing sets. The bars on the top of each figure represent objects. To create these plots a sample of 10,000 objects was used.} 
\label{histogram2}
\end{minipage}
\end{figure*}

\begin{figure*}
\begin{minipage}{150mm}
\begin{subfigure}{.5\textwidth}
  \centering
  \includegraphics[width=.9\linewidth]{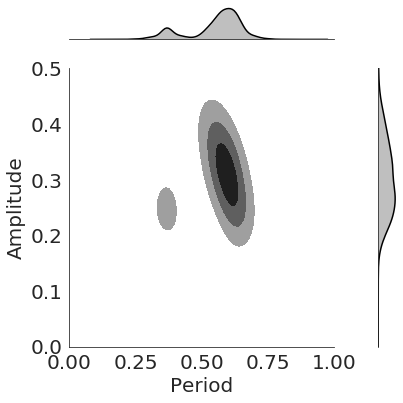}
  \caption{}
  \label{fig:mfig1}
\end{subfigure}%
\begin{subfigure}{.5\textwidth}
  \centering
  \includegraphics[width=.9\linewidth]{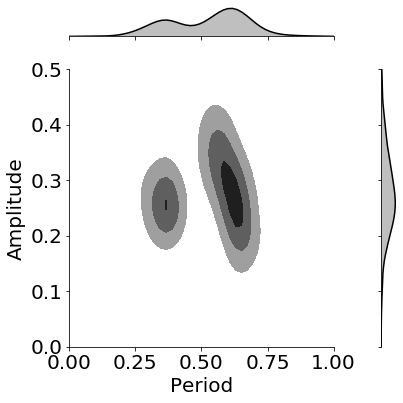}
  \caption{}
  \label{fig:mfig2}
\end{subfigure}
\begin{subfigure}{.5\textwidth}
  \centering
  \includegraphics[width=.9\linewidth]{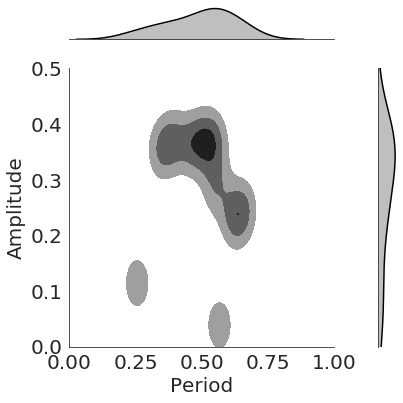}
  \caption{}
  \label{fig:mfig3}
\end{subfigure}%
\begin{subfigure}{.5\textwidth}
  \centering
  \includegraphics[width=.9\linewidth]{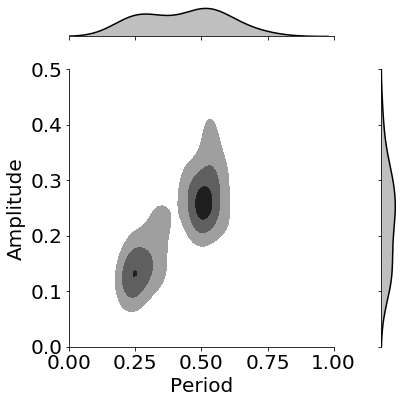}
  \caption{}
  \label{fig:mfig4}
\end{subfigure}
\caption{Density plots for RR Lyrae variable stars in rrlyrae-1 dataset. (a) Small Magellanic Cloud - Training. (b) Small Magellanic Cloud - Testing. (c) Galactic disk - Training. (d) Galactic disk - Testing.}
\label{density}
\end{minipage}
\end{figure*} 

This section presents the inputs used to validate our methodology. Section \ref{ogle} describes the OGLE (Optical Gravitational Lensing Experiment) catalogue. Section \ref{designmatrix} describes how we obtain the final training set from the raw light curves. In Section \ref{biasedSamples}, the procedure to obtain a ground truth is explained. Lastly, in Section \ref{classifier}, we present a set of models which are assessed through our method. 

\subsection{OGLE-III Catalog of Variable Stars}
\label{ogle}

For testing purposes, we use the OGLE-III variable star catalogue, which corresponds to the third phase of the OGLE project \citep{udalski2008optical}. The main goal of OGLE has been to identify microlensing events and transiting planets in four fields: the Galactic bulge, the Large and Small Magellanic Clouds, and the constellation of Carina. We use light curves with at least 25 observations in the $I$ band. The final number of labelled light curves is 420,126. In Table \ref{table1}, we present the number of objects per class.

To estimate the informative priors (step 1 in Figure \ref{procedure}),  we also use the OGLE-III catalogue. When applying the DRs to this dataset, we obtained a subset with $\sim$75\% of RR Lyraes, $\sim$20\% of eclipsing binaries, and $\sim 5\%$ distributed amongst the remaining classes.

\subsection{Processing of Light Curves}
\label{designmatrix}

To extract features from the light curves, we use the Feature Analysis for Time Series (FATS) library \citep{nun2015fats}, thus obtaining a 420,126$\times$63 matrix, where 63 stands for the number of features included in our analysis. Subsequently, to manage both the high dimensionality and multicollinearity, we apply principal component analysis (PCA). \citet{spyroglou2018bayesian} used a similar strategy that combines BLR and PCA to avoid multicollinearity among features. 

\subsection{Shifted training and testing sets}

To evaluate the performance of our approach, we simulate some challenging scenarios, where the training objects are shifted from the testing objects. To create this scenario, we propose a procedure (Algorithm~2) for splitting a labelled catalogue (OGLE-III in our case) into two shifted (biased) datasets. 

\label{biasedSamples}
\begin{center}
\begin{minipage}{1\linewidth}
\begin{algorithm}[H]\small   

\begin{algorithmic}
\renewcommand{\algorithmicrequire}{\textbf{Input:}}
\renewcommand{\algorithmicensure}{\textbf{Output:}}
\REQUIRE Data $\mathcal{D} = (\mathcal{D_X}$,$\mathcal{D_Y})$, classifier $m$, bias control ($T$)
\ENSURE Biased Data $\hat{\mathcal{D}} =(\mathcal{D}_{\text{train}},\mathcal{D}_{\text{test}})$
\STATE 
\STATE $m$.fit($\mathcal{D_X}$,$\mathcal{D_Y}$)
\FOR{$(d_x, d_y) \in (\mathcal{D_X},\mathcal{D_Y})$}
\STATE  $P_A, P_B \leftarrow m$.softPredict($d_x$) 
\STATE $h = 1- (P_A^{2}+P_B^{2})$
\STATE $p = e^{-h/T}$
\STATE $r = \text{uniform}(0,1)$.sample()
    \IF {$r\leq p$}
        \STATE $\mathcal{D}_{\text{train}}$.add($(d_x, d_y)$)
    \ELSE  \STATE $\mathcal{D}_{\text{test}}$.add($(d_x, d_y)$)    
    \ENDIF 
\ENDFOR
\RETURN $\hat{\mathcal{D}}$
\end{algorithmic}
\caption{Procedure to introduce bias in the distribution of objects from a catalogue}
\end{algorithm}
\end{minipage}
 
\end{center}

Firstly, we fit a binary classifier ($m$) that is trained with the entire catalogue. We use an RF Classifier ($m$) to obtain a soft prediction (probability) for each star, and then, we use these predictions to split the dataset $\mathcal{D}$. To split the objects, we define a threshold to assess whether an object can be easily classified or not; to measure that, we use the following metric for each object $i \in \mathcal{D}$: $h_i = 1-(P(A)_i^{2}+P(B)_i^{2})$. This is based on the Gini impurity index \citep{raileanu2004theoretical}, where $P(A)_i^{2}$ is a soft prediction for the true class (RR Lyrae) and $P(B)_i^{2} = 1- P(A)_i^{2}$ a prediction for the false class.  

To avoid a hard threshold when deciding the set (training or testing) for each object, we add a random selection, which is tuned by a constant $T$. This is based on the annealing principle \citep{van1987simulated} and allows us to provide a probabilistic selection of objects, assigning difficult objects ($P(A)_i$ close to 0.5) more frequently to the testing set. As higher values for $T$ are defined a less shifted sets is generated.

We apply Algorithm~2 to obtain datasets with different levels of bias for the RR Lyrae class. The bias was managed by the parameter $T$, and we obtained the datasets rrlyrae-1, rrlyrae-2 and rrlyrae-3 for $T \in \{1,2,4\}$. These three configurations allow us to evaluate our proposal under different bias scenarios. Table \ref{tabledataset} provides a summary of different biased datasets.

Figure \ref{histogram2} shows the hardness distribution (classification difficulty) for objects in the training and testing sets. As we said before, we assume that objects whose prediction scores are close to 0.5 are more difficult to classify than whose predictions scores are close to 1 or 0. According to this definition, in the training sets in Figures \ref{histogram2}(a), \ref{histogram2}(c) and \ref{histogram2}(e), we can observe that the training sets have a higher frequency of easier objects than the testing sets in Figures \ref{histogram2}(b), \ref{histogram2}(d) and \ref{histogram2}(f). The relative frequency of objects at different levels of hardness can be visualized in both type plots, in the histograms and those bars on the top of each figure. 
 
Figure \ref{density} presents the resulting amplitude vs. period distribution, also known as {\it Bailey diagram}, obtained using Algorithm~2, in the space of features for dataset rrlyrae-1. Figures \ref{density}(a) and \ref{density}(b) show a clear shift in the joint distribution of period and amplitude for RR Lyrae from the Small Magellanic Cloud between the training and test sets. Figures \ref{density}(c) and \ref{density}(d) show a similar behaviour for RR Lyrae of the Galactic disk. We note that the bimodal distributions that are seen in these Bailey diagrams are similar to those typically found for RR Lyrae stars \citep[e.g.,][and references therein]{catelan2014pulsating}. In particular, stars in the sequence with the longest periods at any given amplitude are fundamental-mode pulsators, also known as ab-type RR Lyrae stars. Conversely, stars located in the sequence with relative short periods and small amplitudes are first-overtone pulsators, or c-type RR Lyrae stars (RRc). Double-mode RR Lyrae, which pulsate simultaneously in the fundamental and first-overtone modes, also exist, and are commonly denoted as RRd. Their position in the Bailey diagram will depend on which mode is selected as the dominant one.   We note that c-type and d-type RR Lyraes stars are mainly assigned to testing sets (see Figures \ref{density}(b) and \ref{density}(d)). In other words, when we trained the RR Lyrae classifier ($m$) in Algorithm~2 these types (RRc and RRd) were more difficult to classify.

\begin{table*}
\centering
\caption{Number of objects in the training and testing sets for each class. TC represents the true class.}
\begin{tabular}{ccccc}
\hline
$\mathcal{D}$ &  Training &  Testing & TC training & TC testing \\
  \hline
 rrlyrae-1 & 389,364 & 30,762 & 27,240 (6.9\%) & 15,269 (49.6\%)\\ 
 rrlyrae-2 &  402,787 &  17,339 & 34,233 (8.5\%) & 8,500 (49.0\%)\\
 rrlyrae-3 &  335,721 & 84,405 &  34,001(10.1\%) &  8,732(10.3\%) \\
  \hline
\end{tabular}
\label{tabledataset}
\end{table*}

\subsection{Classifiers}
\label{classifier}

As mentioned before, we focus on assessing and ranking a set of BLR classifiers. We compare rankings provided by our method with the Accuracy-based rankings in a CV framework, considering two traditional logistic regression variants. Below we present a brief description of each of these models.

\noindent \textbf{Standard Logistic Regression (LR):} The standard LR classifier models the success probability of a binary dependent variable, $y\in \{0,1 \}$, by means of a Bernoulli distribution,  
    \begin{align}
            p(y|\mathbf{x},\theta) = \text{Ber}(y | s(\mathbf{x},\theta)).
    \end{align}
   \noindent In this model, a sigmoid function,    
    \begin{align}
          s(\mathbf{x},\theta) = \dfrac{1}{1+e^{-\theta^{T}\mathbf{x}}} = \dfrac{e^{\theta^{T}\mathbf{x}}}{1+e^{\theta^{T}\mathbf{x}}},
    \end{align}

    \noindent of input ($\mathbf{X}$) and parameters ($\theta$) is used to model the Bernoulli parameter ($p = s(\mathbf{x},\theta)$).  The Likelihood function,   

\begin{align}
          p(y|\mathbf{x},\theta) = s(\mathbf{x},\theta)^{y}(1-s(\mathbf{x},\theta))^{1-y},
    \end{align}
\noindent is optimized, giving rise to the Maximum Likelihood estimator.

 \noindent \textbf{Penalized Logistic Regression ($l_2$-LR-C):} 
In Bayesian terms, penalized LRs ($l_2$ and $l_1$) embody a prior distribution over $\theta$, and subsequently, the maximum value for the resulting distribution (Maximum a posteriori or MAP) is selected. In particular, $l_2$-LR   is equivalent to a vague Gaussian prior centred at the origin. Let $1/C$ be the penalization factor; hence, if $C$ is small, we obtain a stronger regularization. This approach does not use human knowledge to define the shape of priors. 

\noindent \textbf{Bayesian Logistic Regression (BLR):} BLR focuses on estimating and using the posterior of the distribution of the weights $p(\theta|\mathcal{D})$ in the LR. In our proposal, the informative priors  $\sim \mathcal{N}(\hat{\theta}, \hat{\sigma})$ are estimated using the method laid out in Section \ref{marginallikehood}. For these experiments, we consider DRs for period and amplitude, both of which were estimated with the FATS library \cite{nun2015fats}. 

Each of these models (LR, $l_2$-LR and BLR) represents a family of models ($\mathcal{M}$), which are defined by transformations over their input matrices. Regarding these transformations, first, we apply a linear transformation using PCA retaining the $r$ most important principal components, where $r\in \{ 2, 4, 6, 8, 10, 12\}$, and after that, we also apply polynomial transformations over each component $p \in \{1, 2\}$. The interactions among the components were not considered. 

Let ($\mathbf{X}^{n\times q}$) be the final matrix, where $n$ are the objects in the training set and $q$ the product between the polynomial degree and the number of components used in each model, $m \in \mathcal{M}$. This processing allows us to control the complexity of the model by increasing the number of PCA components or by increasing the degree of the polynomial transformation. Due to convergence problems, the model $m(2,1)$ was not considered in our experiments ($|\mathcal{M}|=11$). Downsampled data were used to deal with imbalanced classes.

In LR and  $l_2$-LR, the models are sorted by their cross-validated Accuracy in training. The BLR models are ordered according to our method (informative marginal likelihood). In the following section, we compare the rankings obtained and show empirical results in which the marginal likelihood (BLR case) can provide improved rankings with respect to CV (LR and $l_2$-LR cases).

\section{Implementation}
\label{implementation}
Our methodology was implemented using Python 3.7. The most important libraries in our code are presented below:
\begin{itemize}
    \item[] \textbf{Pymc3:} probabilistic modelling framework and posterior sampling algorithms \citep{salvatier2016probabilistic}.
    \item[] \textbf{Scikit-Learn:} preprocessing, traditional machine learning models (e.g., Logistic Regression, RF and PCA), cross-validation methods and metrics for assessing models \citep{pedregosa2011scikit}. 
    \item[] \textbf{Pandas:} methods for reading and managing datasets \citep{mckinney2011pandas}.
    \item[] \textbf{Seaborn:}  visualization methods, e.g., scatter plots, histograms and density plots \citep{waskom2014seaborn}.  
\end{itemize}
 We also use a Python implementation of bridge sampling, which was developed by \citet{grunwald2004tutorial}. Lastly, the code source is available at \href{https://github.com/frperezgalarce/vsbms}{https://github.com/frperezgalarce/vsbms}.

\section{Results}
\label{results}

\begin{figure*}
\begin{minipage}{140mm}
\begin{subfigure}{1\textwidth}
  \centering
  \includegraphics[width=0.9\linewidth]{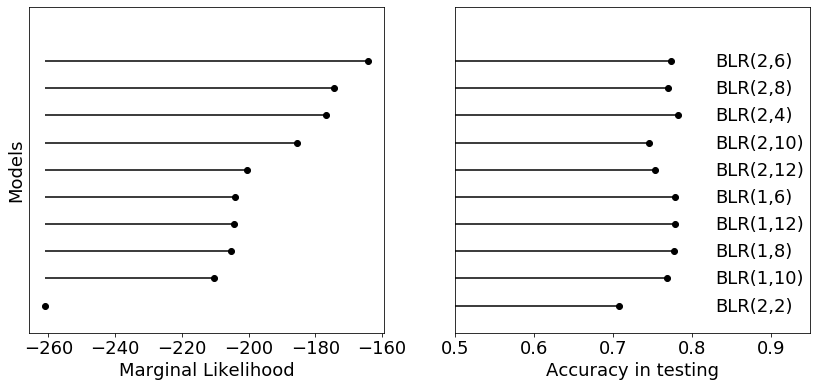}
  \caption{}
  \includegraphics[width=0.9\linewidth]{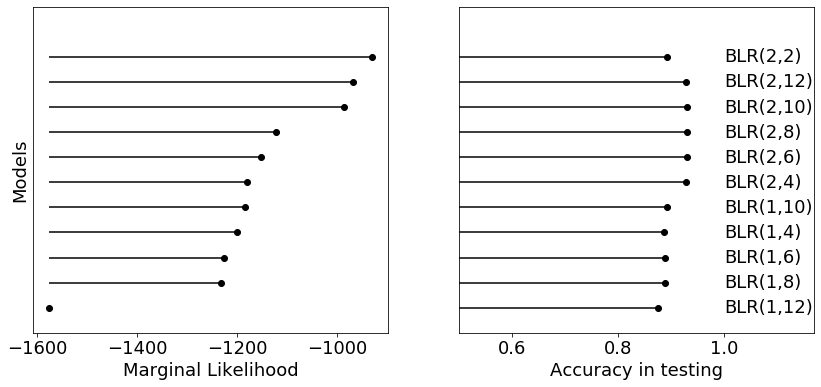}
  \caption{}
  \includegraphics[width=0.9\linewidth]{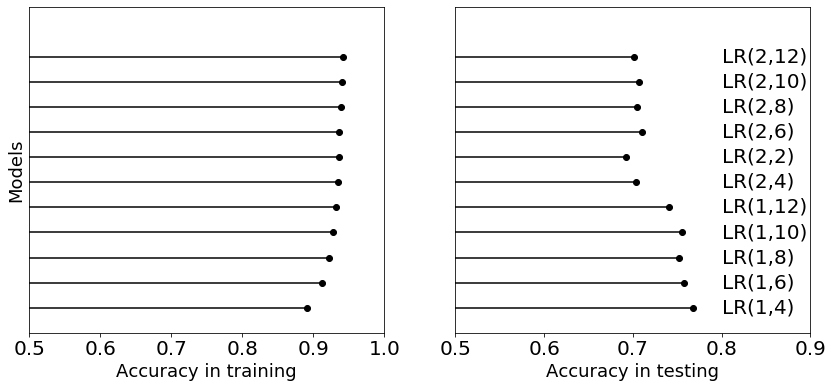}
  \caption{}
  \label{fig:nfig2}
\end{subfigure}
\caption{ Comparison of model rankings with 1,000 samples on rrlyrae-3 set. (a) Models sorted by the marginal likelihood \textbf{BLR-IP}. (b) Models sorted by the marginal likelihood \textbf{BLR-IP}($\sigma$=10). (c) Models sorted by a cross-validated ($k$=10) Accuracy for the $l_2$-\textbf{LR}-1 family of models. Let \textbf{BLR}($p$,$c$) and \textbf{LR}($p$,$c$) be classifiers which is defined by an $n\times(p\times c)$ input matrix where $r$ represents the retained principal components and $p$ the polynomial transformation degree.} 
\label{ranking}
\end{minipage}
\end{figure*}

Figure \ref{ranking} presents two examples of rankings that were generated by different strategies for selecting models: Figure \ref{ranking}(a) provides a ranking of models from our proposed method, while Figure \ref{ranking}(c) shows a ranking using a $k$-fold cross-validated ($k = 10$) Accuracy. In these simple examples, we can note that the marginal likelihood strategies (Figures \ref{ranking}(a) and \ref{ranking}(b)) provide a better ranking coherence compared to the cross-validated Accuracy. In fact, according to Figure \ref{ranking}(c), the cross-validated Accuracy selects the worst model.  

To obtain a more rigorous comparison of rankings among methods, we define a set of metrics to quantify several viewpoints thereof. The selection of metrics used to compare models is presented below. 

\begin{itemize}
    \item \textbf{Kendall-tau ($\tau$):} this metric is estimated by $ \dfrac{n_c-n_d}{\frac{1}{2}n(n-1)}$, where $n_c$ represents the number of concordant models and $n_d$ is the number of discordant models in the ranking. It identifies the coincidences between training and testing rankings. Both rankings (training and testing) are concordant when the selection  is correct. In preliminary experiments,  similar results were obtained by using the Spearman's rank coefficient.    
    \item \textbf{Average top-3-Accuracy ($A$):} In order to discriminate beyond the rankings, we also use the average Accuracy (in test) over the three best models for each ranking. Thus, we can identify the quality of the selected models. This metric provides a perspective about how good are the models prioritized by each strategy. Note that the best model Accuracy can be a harsh metric in our context (ranking of models). On the other hand, the average performance over a family of models is a poorly informative measure. 
    \item \textbf{Average top-3-$F_1$-score ($F_1$):} As was explained in Section \ref{confusionmatrix}, this metric is a better option than $A$ in case of having unbalanced classes. To assess the performance of each method, we estimate the average $F_1$-score in testing for the three foremost models in the ranking. 
    \item \textbf{Delta training/testing ($\Delta_T$):} This metric seeks to evaluate how far, on average, the predicted Accuracy for the testing set is with respect to the training set. We report the average distance over the set of models in each family.   
\end{itemize}

In Tables \ref{table10}-\ref{table12}, we summarize the metrics for each dataset (rrlyrae-1, rrlyrae-2 and rrlyrae-3), considering three subset sizes $s$ (for training). As was commented in Section \ref{classifier}, we run three baseline strategies to rank models in addition to three approaches based on the marginal likelihood. The baseline approaches (LR, $l_2$-LR-$1$ and $l_2$-LR-$100$) are based on $k$-fold CV with $k = 10$. The rankings based on the marginal likelihood consider flat priors (BLR-FP); informative priors for the mean and fixed variance (BLR-IP $\sigma = 10$); and informative priors on both the mean and variance (BLR-IP).

\begin{table}

\centering
\caption{Evaluation of rankings of models in rrlyrae-1. $\tau$ is the Kendall's tau rank correlation; A and F1 are the mean Accuracy and the mean F1-score, respectively; of the three foremost models, $\Delta_T$ is the average difference between the Accuracy in training and testing. The bold numbers represent the best strategy for model selection by each metric.
}
\begin{tabular}{rr|cccc}
\hline
 \multicolumn{6}{c}{$k$-fold CV}\\
\hline
$\mathcal{M}$ &        $s$ &         $\tau$ &     $F_1$    & A &   $\Delta_{T}$  \\ 
\hline 
 LR &   1000 &  0.11 &  0.62 &  0.62 &  0.35 \\
 &  2000 &  0.09 &  0.62 &  0.63 &  0.34 \\
 &  4000 &  0.26 &  0.63 &  0.65 &  0.32  \\
\hline
$l_2$-LR-100& 1000 & -0.31 &  0.64 &  0.64 &  0.35 \\
            &  2000 & -0.09 &  0.64 &  0.63 &  0.35 \\
            &  4000 &  0.24 &  0.68 &  0.66 &  0.32 \\

\hline
$l_2$-LR-1 &    1000 & -0.16 &  0.64 &  0.64 &  0.35 \\
           &  2000 & -0.02 &  0.63 &  0.63 &  0.34 \\
           &  4000 &  0.49 &  0.69 &  0.66 &  0.32  \\
\hline
 \multicolumn{6}{c}{marginal likelihood}\\
\hline
$\mathcal{M}$ &        $s$ &         $\tau$ &     $F_1$    & A &   $\Delta_{T}$ \\
\hline 
BLR-FP&    1000 & 0.70 &  \textbf{0.70} &  \textbf{0.69} &  0.32  \\
   &   2000 &  0.82 &  \textbf{0.70} &  \textbf{0.69} &  0.32 \\
   &  4000 &  \textbf{0.85} &  \textbf{0.70} &  \textbf{0.69} &  0.31 \\
\hline
BLR-IP ($\sigma$=10) &   1000 &  0.31 &  0.68 &  \textbf{0.69} &  0.32  \\
&  2000 &  0.56 & \textbf{0.70} &  \textbf{0.69} &  0.32  \\
&  4000 &  0.75 &  \textbf{0.70} &  \textbf{0.69} &  0.31 \\
\hline
BLR-IP  &    1000 &  0.60 &  0.50 &  0.61 &  \textbf{0.24}\\
        &  2000 &  \textbf{0.85} &  0.65 &  0.66 &  0.34 \\
        &  4000 &  0.71 &  0.69 &  0.68 &  0.32  \\
\hline
\end{tabular}
\label{table10}

\end{table}

\begin{table}
\footnotesize
\centering
\caption{As in Table \ref{table10}, but for the case of rrlyrae-2.
}
\begin{tabular}{rr|cccc}
\hline
 \multicolumn{6}{c}{$k$-fold CV}\\
\hline
$\mathcal{M}$ &        $s$ &         $\tau$ &     $F_1$    & A &   $\Delta_{T}$ \\
\hline 
 LR &    1000 &  0.13 &  0.67 &  0.64 &  0.32 \\
    &  2000 &  0.38 &  0.66 &  0.66 &  0.32 \\
    &  4000 &  0.38 &  0.63 &  0.65 &  0.32\\
\hline
$l_2$-LR-100 & 1000 &  0.27 &  0.65 &  0.64 &  0.32 \\
             &  2000 &  0.24 &  0.65 &  0.63 &  0.33 \\
             &  4000 &  0.45 &  0.65 &  0.63 &  0.34 \\
\hline
$l_2$-LR-1  & 1000 &  0.42 &  0.66 &  0.65 &  0.32 \\
            &  2000 &  0.27 &  0.65 &  0.63 &  0.33 \\
            &  4000 &  0.45 &  0.65 &  0.63 &  0.34 \\
 \hline
 \multicolumn{6}{c}{marginal likelihood}\\
 \hline
 $\mathcal{M}$ &        $s$ &         $\tau$ &     $F_1$    & A &   $\Delta_{T}$\\
\hline 
BLR-FP&    1000 &  0.71 &  0.68 &  0.66 &  0.32 \\
   &   2000 &  0.64 &  0.68 &  0.66 &  0.33 \\
   &  4000 &  0.76 &  \textbf{0.69} &  \textbf{0.67} &  0.32 \\
\hline
BLR-IP ($\sigma$=10) &    1000 &  0.20 &  0.68 &  \textbf{0.67} &  0.33 \\
                     &  2000 &  0.61 &  0.68 &  0.66 &  0.33 \\
                     &  4000 &  0.73 &  0.68 &  \textbf{0.67} &  0.32 \\
\hline
BLR-IP&    1000 &  0.60 &  0.49 &  0.59 &  \textbf{0.24} \\
&  2000 &  \textbf{0.82} &  0.68 &  0.66 &  0.33 \\
&  4000 &  0.78 &  0.68 &  \textbf{0.67} &  0.33 \\
\hline
\end{tabular}

\label{table11}

\end{table}

\begin{table}

\footnotesize
\centering
\caption{As in Table \ref{table10}, but for the case of rrlyrae-3.}
\begin{tabular}{rr|cccc}
\hline
 \multicolumn{6}{c}{$k$-fold CV}\\
\hline
$\mathcal{M}$ &        $s$ &         $\tau$ &     $F_1$    & A &   $\Delta_{T}$ \\

\hline
 LR &    1000 & -0.75 &  0.46 &  0.76 &  0.19 \\
    &  2000 & -0.53 &  0.45 &  0.75 &  0.20 \\
    &  4000 & -0.05 &  0.57 &  0.85 &  0.12 \\
\hline
$l_2$-LR-100&1000 & -0.75 &  0.46 &  0.76 &  0.20 \\
                  &  2000 & -0.78 &  0.44 &  0.74 &  0.22 \\
                  &  4000 & -0.42 &  0.51 &  0.80 &  0.16 \\
\hline
$l_2$-LR-1 &    1000 & -0.71 &  0.46 &  0.76 &  0.20 \\
           &  2000 & -0.78 &  0.44 &  0.74 &  0.22 \\
           &  4000 & -0.49 &  0.51 &  0.80 &  0.17 \\
\hline
 \multicolumn{6}{c}{marginal likelihood}\\
 \hline
 $\mathcal{M}$ &        $s$ &         $\tau$ &     $F_1$    & A &   $\Delta_{T}$\\
\hline 
BLR-FP&    1000 & -0.16 &  0.49 &  0.78 &  0.19 \\
   & 2000 & -0.36 &  0.46 &  0.76 &  0.22 \\
   &  4000 & -0.39 &  0.46 &  0.76 &  0.21 \\
\hline
BLR-IP ($\sigma$=10)&    1000 &  0.09 &  0.49 &  0.79 &  0.18 \\
&  2000 & -0.30 &  0.47 &  0.76 &  0.22 \\
&  4000 & -0.33 &  0.46 &  0.76 &  0.21 \\

\hline

BLR-IP&   1000 &  \textbf{0.56} &  \textbf{0.71} &  \textbf{0.93} & \textbf{-0.25} \\
&  2000 & -0.53 &  0.47 &  0.76 &  0.22 \\
&  4000 & -0.53 &  0.46 &  0.76 &  0.22 \\
\hline
\end{tabular}
\label{table12}

\end{table}

From Tables \ref{table10}-\ref{table12}, we can also observe that $\tau$ values are greater for marginal likelihood-based rankings than those based on $k$-fold cross-validation.  These results demonstrate empirically that the marginal likelihood is more robust than the cross-validated $A$ for assessing and prioritizing RR Lyrae star classifiers under different levels of bias. In fact, according to $\tau$ values, when looking at rrlyrae-1 and rrlyrae-2 results, the best rankings were provided by BLR-IP (2000), while the best ranking for rrlyrae-3 was obtained using BLR-IP (1000).  

Concerning $F_1$-score and $A$ for the three sets, the best three informative Bayesian models obtain a better performance than the best three likelihood-based models. This means that the predictive performance of posterior samples (posterior mean) is also improved when we add prior knowledge from DRs. When looking at experiments with rrlyrae-1, it is worth noting that the difference is more significant in the $A$ metric. 

When looking the impact of bias (managed by $T$ in Algorithm~2) on $A$ and $\Delta_T$, we can note that in highly biased data sets (rrlyrae-1 and rrlyrae-2) our proposal is the best alternative, but even BMS is competitive in the less biased set (rrlyrae-3).

Table \ref{table12} shows that, given a smaller bias, the six alternatives obtain better $A$. However, in this set, this metric is poorly informative due to the imbalance problem in this testing set (10\% of RR Lyraes see Table \ref{tabledataset}). In spite of that, it also shows a better F1-score value for those models selected by the informative marginal likelihood (BLR-IP). Figure \ref{ranking}(a), \ref{ranking}(b) and \ref{ranking}(c) show the rankings provided by \textbf{BLR-IP} ($\sigma$=10), \textbf{BLR-IP} and $l_2$-\textbf{LR}-1 for this experiment (dataset rrlyrae-3 and $s=1 000$). When looking at Figure \ref{ranking}(c), we can observe that the cross-validated Accuracy is unable to prioritize models correctly; in fact, this approach selects the worst model in this case.

To sum up, when there are biased labelled objects and we have expert knowledge, our scheme provides an excellent alternative to select models. Note that BLR-IP ($s$ = 1000) obtains the best $\Delta_T$ in the three sets; this means that the reported $A$ is more reliable. It comes from informative priors on small training sets can penalize the performance highly in training since this expert knowledge helps to limit the likelihood function (based on data), but when we use them, the selected models are more robust to biases. 

\section{Conclusions}
 \label{label}
We have presented a novel approach to assess and sort models considering expert knowledge. The method is based on the design of informative priors using deterministic physical rules that allow us to estimate an informative marginal likelihood, which includes well-known properties like a model selector. The method offers a good alternative for selecting variable star classifier with biased (and/or small) sets of labelled objects. This gives rise to an original and simple methodology to add prior knowledge in the model assessment process of RR Lyrae star classifiers without having to undergo a time-consuming adaptation process.     

For evaluation purposes, we have designed a method capable of introducing bias to a dataset according to the classification difficulty of each object. This allows us to test different strategies to assess models under three conditions of bias. The results show that the informative marginal likelihood is able to identify more suitable models than non-informative cross-validated metrics. 

Future work can consider extensions such as: (i) the use of other types of informative priors, e.g., priors over the proportion of classes or heavy-tailed distributions over BLR's weights; (ii) analysis of other time-series survey datasets; (iii) and the application of this approach to other classes (or subtypes) of variable stars.  

\section*{ACKNOWLEDGEMENTS}
We acknowledge the support from CONICYT-Chile,
through the FONDECYT Regular project number 1180054. F. P\'erez-Galarce acknowledges the support from National Agency for Research and Development (ANID), through Scholarship Program/Doctorado Nacional/2017-21171036. P. Huijse acknowledges financial support from ANID through project PAI 79170017. Support for M. Catelan is provided by ANID's Millennium Science Initiative through grant ICN12\textunderscore 120009, awarded to the Millennium Institute of Astrophysics (MAS); by Proyecto Basal AFB-170002; and by FONDECYT grant \#1171273.

\bibliographystyle{mnras}
\bibliography{references}

\appendix
\section{Proof of Bridge Sampling Estimator}
\label{bridgeSampling}

The proposed bridge sampling approach \citep{gronau2017tutorial} begins with the following identity: 

\begin{equation}
1 = \frac{\int p(\mathcal{D|\theta})p(\theta)h(\theta)g(\theta)d\theta}{\int p(\mathcal{D|\theta})p(\theta)h(\theta)g(\theta) d\theta}, \label{eq11}
\end{equation}

\noindent where $g(\theta)$ is the proposal distribution. Subsequently, it is multiplied by the marginal likelihood on both sides. Then, we obtain the  following equation: 

\begin{equation}
p(\mathcal{D}) = \frac{\int p(\mathcal{D|\theta})p(\theta)h(\theta)g(\theta) d \theta }{\int \frac{p(\mathcal{D|\theta})p(\theta)}{p(\mathcal{D})}h(\theta)g(\theta) d \theta}. \label{eq12}
\end{equation}
\noindent  Note that the posterior distribution appears on the right side's denominator. After that, by means of 

\begin{equation}
p(\mathcal{D}) = \frac{\int p(\mathcal{D|\theta})p(\theta)h(\theta)
}{\int h(\theta)g(\theta) }\frac{g(\theta) d \theta }{p(\mathcal{\theta | D})d \theta}, \label{eq13}
\end{equation}

\noindent the right side of \ref{eq12} is separated into two ratios, and consequently, we can obtain the expected values in the denominator and numerator as follows: 

\begin{equation}
p(\mathcal{D}) = \frac{\mathbb{E}_{g(\theta)}\left[  p(\mathcal{D|\theta})p(\theta)h(\theta)  \right]}{ \mathbb{E}_{p(\theta | \mathcal{D})} \left[  h(\theta)g(\theta)    \right]}.\label{eq14}
\end{equation}

Finally, we use the definition of optimal bridge function provided by \citet{meng1996simulating}:
\begin{equation}
\label{bridgeFunction_appendix} 
h(\theta) = C\frac{1}{s_1 p(\mathcal{D}|\theta)p(\theta) + s_2p(\mathcal{D})g(\theta)  }.
\end{equation}

Due to the obtained estimator depends recursively on the marginal likelihood, the iterative scheme presented below is applied:
\begin{equation}
\hat{p}(\mathcal{D})^{t+1} = \frac{\frac{1}{N_2} \sum_{i = 1}^{N_2}\frac{p(\mathcal{D}| \theta_i) p(\theta_i)}{s_1 p(\mathcal{D}|\theta_i)p(\theta_i) + s_2\hat{p}(\mathcal{D})^{t}g_(\theta_i) }}{\frac{1}{N_1} \sum_{j = 1}^{N_1}\frac{g(\theta_j)}{s_1 p(\mathcal{D}|\theta_j)p(\theta_j) + s_2\hat{p}(\mathcal{D})^{t}g_(\theta_j) }}.\label{eq16}
\end{equation}
\end{document}